\documentclass{article}
\usepackage{amssymb}
\usepackage{amsmath}
\usepackage{amsfonts}
\usepackage{graphics,epsfig}

\newtheorem{theorem}{Theorem}

\newtheorem{definition}[theorem]{Definition}
\newtheorem{lemma}[theorem]{Lemma}

\newenvironment{proof}[1][Proof]{\noindent\textbf{#1.} }{\ \rule{0.5em}{0.5em}}

\begin{document}

\title{Quantum and classical probability as Bayes-optimal observation.}
\author{Sven Aerts \\ \\
\textit{Center Leo Apostel for Interdisciplinary Studies (CLEA)} \\
\textit{and Foundations of the Exact Sciences (FUND) } \\
\textit{Department of Mathematics, Vrije Universiteit Brussel} \\
\textit{Pleinlaan 2, 1050 Brussels, Belgium} \\
email: \textsf{saerts@vub.ac.be}}

\date{}

\maketitle

\begin{abstract}
We propose a simple abstract formalisation of the act of observation, in which the
system and the observer are assumed to be in a pure state and their
interaction deterministically changes the states such that the outcome can
be read from the state of the observer after the interaction. If the observer
consistently realizes the outcome which maximizes the likelihood
ratio that the outcome pertains to the system under study (and not to his
own state), he will be called Bayes-optimal. We calculate the probability if
for each trial of the experiment the observer is in a new state
picked randomly from his set of states, and the system under investigation
is taken from an ensemble of identical pure states. For classical
statistical mixtures, the relative frequency resulting from the maximum
likelihood principle is an unbiased estimator of the components of the
mixture.  For repeated Bayes-optimal observation in case the state space is
complex Hilbert space, the relative frequency converges to the Born rule. 
Hence, the principle of Bayes-optimal observation can be regarded as an underlying mechanism for the Born rule. 
We show the outcome assignment of the Bayes-optimal observer is invariant under unitary transformations and contextual, 
but the probability that results from repeated application is non-contextual. 
The proposal gives a concise interpretation for the meaning of the occurrence of a single outcome in a quantum
experiment as the unique outcome that, relative to the state of the system,
is least dependent on the state of the observe at the instant of measurement.
\end{abstract}

\section{Introduction}

As early as 1935, Schr\"{o}dinger wrote:\emph{\ ``The
rejection of realism has logical consequences. In general, a variable has no
definite value before I measure it; then measuring it does not mean
ascertaining the value that it has. But then what does it mean?"} \cite{Schrodinger}. As the
advent of quantum mechanics solved the long standing problem of providing an
adequate description for several important and unexplained experiments, the
problem of realism in quantum mechanics was initially perceived mainly as a
challenge to the construction of a new philosophy of natural science. In
support of this perception, is the fact that almost all later theoretical
advances with experimental consequences came about without any serious
progress with this very basic problem. Yet at the same time, a growing
number of people recognized that progress in this problem would likely have
deep consequences for the quantum-classical transition, the attempt to
produce a successful unification of quantum mechanics and relativity theory,
and the related problem of quantum cosmology. Halfway the sixties two
important advances were made. In 1964, John Bell showed that any local
hidden variable theory will yield predictions that are at odds with quantum
mechanics. A few years later, Kochen and Specker \cite{KochenSpecker}
presented an explicit set of measurements, for which the simultaneous
attribution of values for each of these measurements, leads to a logical
contradiction. The two results can be regarded as opposite faces of the same
coin. Whereas Bell's result can be verified (or refuted) by experiment,
Kochen and Specker's argument shows the problem also to be a deeply-rooted
theoretical one. These two results have been of such importance, that the
notion of realism in quantum physics is usually considered automatically as
having either the meaning of `locally realistic' (Bell), or that of `the
impossibility of attributing predetermined outcome values to the set of
observables' (Kochen and Specker). The apparent lack of realism in quantum
mechanics has been illustrated again and again by clever theoretical
constructions ranging from Bell-type arguments to impossible coloring games,
and the countless attempts to produce an as loophole free as possible
experimental verification of these arguments \footnote{%
Because local theories, by Bell's theorem, cannot give rise to some of the
experimentally verifiable predictions of quantum mechanics, the requirement
of locality, or so-called ``local-realism" takes a prominent role. However,
realism seems more fundamental than locality, in the sense that the latter
is only well-defined if we can attribute some form of reality with
respect to the whereabouts of the system. Moreover, the derivation of the
quantum correlation for most Bell-type experiments do not, at any point,
invoke spatial coordinates. As far as concerns the actual application of
quantum theory, it is quite immaterial whether we calculate the correlations
between various outcomes that are obtained in a single location or at
space-like separated locations. Of course, for a locally realistic theory,
the difference is huge.}.

However, the commonly accepted notion that \textquotedblleft measuring a variable does not mean ascertaining the
value that it has\textquotedblright, does not mean that the answer to 
Schr\"{o}dinger's question is that the occurrence of a particular outcome has 
\emph{no} meaning. Every proper quantum experiment is a testimony to the
contrary, for if a single outcome has no informational content about the
system at all, then how are we to derive anything at all from the sum of a
great number of informationally empty statements? Whether we perform a
tomographic state reconstruction, or experimentally estimate the value of a
physical quantity of a system, we accept that in a well constructed
experiment every outcome presents a piece of information, a piece of
evidence, that brings us closer to the true state of affairs, whatever that
may be. To give a more detailed answer to the question, we are in need of a
model that shows \emph{how} a single outcome is obtained. We will provide
such a model in an attempt to understand the meaning of the occurrence of a
single outcome in a quantum mechanical experiment. More specifically, we
will show that an observer actively seeking to minimize his own influence on
the produced outcome, will, with the aid of Bayesian decision theory, give outcomes
whose relative frequency converges to the Born rule in a natural way. This
in turn will give us a possible interpretation for the occurrence of a
particular outcome.

\section{Probabilities of outcomes for a single observable quantity}

Let us assume we have a system $S$ for which we write $\Sigma _{S}$ to denote its set of states, 
and $\mathcal{A}$ for an observable that can take any single outcome out of $n$ distinct values in the outcome set 
$X=\{x_{1},\ldots ,x_{n}\}$. At the most trivial level, there is a counting
measure on the set of outcomes. If $\mathcal{P}(X)$ denotes the set of all
subsets of $X$, then the probability that a measurement of observable $%
\mathcal{A}$ on the system in a state $\psi \in \Sigma _{S}$ yields an
outcome in a given subset $X_{1}\in \mathcal{P}(X),$ is a mapping 
\begin{equation}
p(.|.):\mathcal{P}(X)\times \Sigma _{S}\rightarrow \lbrack 0,1]
\label{prob prescript}
\end{equation}%
such that for disjoint $X_{i}\in $ $\mathcal{P}(X),$ we have: 
\begin{equation}
p(\cup X_{i}|\psi )=\sum_{i}p(X_{i}|\psi )  \label{prob add}
\end{equation}%
The additive property described by (\ref{prob add}) is generally accepted
both in quantum and classical probability and provides the rationale for the
use of normalized states, that is, states $\psi $ that satisfy: 
\begin{equation}
p(X|\psi )=1  \label{prob normalized}
\end{equation}%
In this way, (\ref{prob normalized}) reduces the number of free parameters
in state space by one. We have written $p(x|\psi )$ to emphasize that it
represents the probability that the outcome $x$ obtains when (we know that)
the system is prepared in the state $\psi .$ The classical interpretation
for the arisal of probabilities, is one of a lack-of-knowledge about the
precise state being measured. From a naive epistemic perspective, the
outcome $x$ is then an objective attribute of each measured state, and the
probability related to each outcome is simply the fraction of states \emph{%
having} the \textquotedblleft $x$-attribute\textquotedblright\ in the
ensemble of systems that we measure. As indicated in the introduction, such
an interpretation for the probabilities in quantum mechanics is problematic.
Even for a single spin 1/2 particle, one can show \cite{Accardi} three
measurements suffice to exclude such an interpretation, even without taking
recourse to locality issues.

\subsection{Quantum probability for a single observable quantity}

In orthodox quantum mechanics, the state space $\Sigma _{S}$ is the complex
Hilbert space $\mathcal{H}$. The set of states of the observed system that
we will consider, is the set of unit vectors in an $n$-dimensional Hilbert
space $\mathcal{H}_{n}$,%
\begin{equation}
\Sigma _{S}=\{\psi \in \mathcal{H}_{n}  :|\psi |=1 \}  \label{state space}
\end{equation}%
As usual, the norm $|.|$ is defined through the (sesquilinear) inner product that we will denote $\langle .|.\rangle $. 
Alternatively, one can take rays or even density operators for the states. Since both lead to essentially the same results, we will stick to unit norm vectors.
Let $\mathcal{L(H}_{n}%
\mathcal{)}$ be the set of linear operators that act on the elements of $%
\mathcal{H}_{n}$, then an observable $\mathcal{A}$ is represented by a
self-adjoint element of $\mathcal{L(H}_{n}\mathcal{)}$:%
\begin{equation}
\mathcal{A\in L(H}_{n}\mathcal{)}:\mathcal{A}^{\dagger }\mathcal{=A}
\label{selfadjoint}
\end{equation}%
Throughout this presentation, we assume $\mathcal{A}$ has a discrete,
finite, non-degenerate spectrum, which implies that eigenvectors belonging
to different eigenvalues are orthogonal. Let $F_{\mathcal{A}}$ be the set of
the eigenvectors%
\footnote{Again, we neglect mathematical details with regard to phase issues and identify 
all $\psi$ with the same eigenvalue  $c_i$ as the same eigenvector $\psi_i$.} 
of $\mathcal{A}$%
\begin{equation}
F_{\mathcal{A}}=\{\psi _{i}\in \mathcal{H}_{n}:\mathcal{A}|\psi _{i}\rangle
=c_{i}|\psi _{i}\rangle ,\ c_{i}\in 
\mathbb{R}
\}  \label{frame}
\end{equation}

We now have $\langle \psi _{i}|\psi _{j}\rangle =\delta _{i,j}$ and $%
\sum_{i}|\psi _{i}\rangle \langle \psi _{i}|=I,$ and, because the spectrum is assumed non-degenerate, we have that $F_{\mathcal{A}%
}$ is a basis or a complete orthonormal frame. From linear algebra we know that an arbitrary
element $\psi ^{s}$ of $\mathcal{H}_{n}$ can be written in this frame $F_{%
\mathcal{A}}$ as: 
\begin{equation}
|\psi ^{s}\rangle =\sum_{i=1}^{n}\alpha _{i}|\psi _{i}\rangle
\label{general state}
\end{equation}%
If $\psi ^{s}$ satisfies (\ref{prob normalized}), then it lies in $\Sigma
_{S}\subset \mathcal{H}_{n}$ , and the $\alpha ^{\prime }s$ obey:%
\begin{equation}
\sum_{i}\alpha _{i}\alpha _{i}^{\ast }=1  \label{normalization}
\end{equation}%
Moreover, one can easily verify that the observable $\mathcal{A}$ can be
written as%
\begin{equation}
\mathcal{A=}\sum_{i}a_{i}|\psi _{i}\rangle \langle \psi _{i}|
\label{operator}
\end{equation}

Hence the observable $\mathcal{A}$ is in a one-to-one correspondence with an
orthonormal frame $F_{\mathcal{A}}$ of eigenvectors of $\mathcal{A}$ and we
will represent the observable by its associated frame. Throughout this
paper, we reserve superscripts of states as a mnemotechnical aid for system
recognition (i.e. $\psi ^{s}$ is a system state and $\psi ^{m}$ the state of
the measurement apparatus) and subscripts of states to denote eigenstates.
If a system is in an eigenstate corresponding to outcome $x_{i},$ we will
denote the corresponding eigenstate as $\psi _{i}.$ For an arbitrary
eigenstate $\psi _{j}$ we have%
\begin{equation}
p(x_{i}|\psi _{j})=\delta _{i,j}  \label{eigenstates}
\end{equation}%
Thus for an eigenstate, and also for a statistical mixture of eigenstates,
the classical interpretation of probability as \textquotedblleft proportion
of system having the $x$-attribute\textquotedblright\ is tenable. The more
interesting case, however, is the probability for the occurrence of an
outcome $x_{i}$ when the system is in a general state (\ref{general state}),
which is given by the Born rule: 
\begin{equation}
p(x_{i}|\psi ^{s})=|\langle \psi _{i},\psi ^{s}\rangle |^{2}=|\alpha
_{i}|^{2}  \label{Born}
\end{equation}

The analog with the classical situation would be that $\psi ^{s}$ represents
a mixture of states that have attribute $x_{k}$ in the right proportion such
that the\ Born rule holds. However the Born rule holds even when the system
is in a pure state, i.e. a state which cannot be obtained as a statistical
mixture of states. We will show that it is possible to regard the
probabilities as arising from a lack of knowledge about the detailed state
of the observer if the observer actively attempts to choose the outcome
that maximizes a specific likelihood ratio that we will present shortly.

\section{The process of observation}

\subsection{The deterministic observer}

Let us first define what we mean by an observer. An observer is a physical
system that takes a question as input, and yields in reply an outcome which
is a member of a discrete set. This outcome can be freely copied, and hence
communicated to many other observers. In general, this definition of
observer will include the experimental setup, apparata, sensors, and the
human operator. It is however quite irrelevant to our purposes whether we
consider an apparatus or a detector, an animal or a human being as observer,
as long as we agree that it is this system that has produced the outcome. We
will furthermore assume the observer comes to this outcome through \emph{a
physical, deterministic interaction}. That is, if we have perfect knowledge
of the initial state of the system and of the potentials that act on the
system, we can in principle predict the future state of the system
perfectly. Besides the fact that all fundamental theories of physics (even
classical chaotic systems and quantum dynamics) postulate deterministic
evolution laws, the requirement of determinism allows to derive probability
as a secondary concept. So let us assume that the outcome of an observation
is the result of a deterministic interaction:

\begin{equation}
\tau :\Sigma _{S}\times \Sigma _{M}\rightarrow X  \label{interaction}
\end{equation}

Here $\tau $ is the interaction rule, $\Sigma _{S}$ is the set of states of
the observed system, $\Sigma _{M}$ the set of states of the observing system
and $X$ the set of outcomes that observable $\mathcal{A}$ can have. We will
deal only with a single observable, so no further notational reference is
made to the particular observable. The mapping $\tau $ encodes how an
observer in a state $\psi ^{m}\in \Sigma _{M},$ observing a system in the
state $\psi ^{s}\in \Sigma _{S}$, comes to the outcome $x\in X.$ Because our
observer is deterministic, we assume $\tau $ is single-valued. Probability
will only arise as a lack of knowledge on deterministic events. The observer
faces the task of selecting an outcome from the set $X$ that tells something
about the system under observation. But the outcome is always formulated by
the observer, it has to be encoded somehow in the state of the observer
after the observation. Hence the outcome \emph{itself} is also an observable
quantity of the post-measurement state of the observer. The outcome will
then have to share its story among the two participating systems that gave
rise to its existence: it will always have something to say about both the
observer \emph{and} the system under study. In \cite{Aerts undecidable} it
was shown by a diagonal argument, that even in the most simple case of a
perfect observer, observing only classical properties\footnote{%
We say the property $\mathbf{a}$ of a system $S$ in the state $s$ is actual,
iff the testing of property $\mathbf{a}$ for $S$ in the state $s$, would yield an
affirmation with certainty. A property is called classical when the outcome
of the observation to test that property, was predetermined by the state of
the sytem (whatever that state was) prior to the test. For a classical
property we can define a negation in the lattice of properties that is
simply the Boolean NOT. A property $\mathbf{a}$ is then classical for $S$
iff for each state of $S$ the property, \emph{or} its negation, is actual.
For details, see \cite{Aerts undecidable}.}, there exist classical
properties pertaining to himself that he cannot perfectly observe. More
specifically, even if the observer can observe a given (classical) property
perfectly, he cannot perfectly observe \emph{that} he observes this
classical property perfectly. There is no logical certainty with respect to
faithfulness of a single shot, deterministic observation. On the other hand,
observation is an absolutely indispensable part of doing science, hence it
is only natural that every scientist believes that faithful observation can
and does indeed occur. Living in the real world, somewhere between the
extremes of the ideal and the impossible, we wonder whether there is a
strategy for the observer so that he is guaranteed that each outcome he
picks uses his observational powers to the best of his ability.

\subsection{Repeated measurement and the randomization of probe states of
the observer}

Rather than attempting to measure observables in a single trial of an
experiment, our observer turns to a new strategy. First he prepares an
ensemble of a large number of identical system states. Next he will interact
with each of the members of this ensemble in turn. For each and every single
interaction, he will pick the outcome that somehow `has the largest
likelihood' of pertaining to the system. By randomizing his probe state and
picking the outcomes in this way, the observer hopes to restore objectivity,
so that he will eventually obtain information that pertains solely to the
system under observation. To calculate $p(x|\psi ^{s})$ within the
deterministic setting of the previous section (\ref{interaction}) is in
principle straightforward. The experiment our observer will perform is a
repeated one, in which the set of states of the system under study is
reduced to a singleton, and the set of states for the observer is the whole
of $\Sigma _{M}$. The set of states for the observer that leads to a given
outcome $x\in X$ when the observer observes a system in the state $\psi ^{s}%
\footnote{%
In accordance with the literature on the subject, we used Dirac's bra-ket
notation for our brief introduction to quantum probability. In what follows
we will not make use of the duality between a Hilbert space and the space of
linear functionals on this Hilbert space, so all vectors are written without
brackets.}$ will be denoted as $eig(x,\psi ^{s})$: 
\begin{equation}
eig(x,\psi ^{s})=\{\psi \in \Sigma _{M}:\tau (\psi ^{s},\psi )=x\}
\label{eigensets}
\end{equation}%
From the single-valuedness of $\tau $ in (\ref{interaction}), we have for $%
x_{i}\neq x_{j}$: 
\begin{equation}
eig(x_{i},\psi ^{s})\cap eig(x_{j},\psi ^{s})=0  \label{eigpart 1}
\end{equation}%
If we assume that the act of observation of an observable leads to an
outcome for every state of the system investigated, we have

\begin{equation}
\cup _{i=1}^{n}eig(x_{i},\psi ^{s})=\Sigma _{M}  \label{eigpart 2}
\end{equation}%
In this way $\tau $ defines in a trivial way a \emph{partition} of the state
space of the observer with each member $eig(x_{i},\psi ^{s})$ in the
partition belonging to exactly one outcome. We are now ready to introduce
probability. With $\mathcal{B}(\Sigma _{M})$ a $\sigma$-algebra of Borel
subsets of $\Sigma _{M},$ (which we tacitly assume includes $eig(x_{i},\psi
^{s})$ for every $i$), we define a probability measure $\mu $ that acts on
the measure space $(\Sigma _{M},\mathcal{B}(\Sigma _{M})).$ For any two
disjoint $\sigma _{i},\sigma _{j}$ in $\mathcal{B}(\Sigma _{M})$, we have%
\begin{eqnarray}
\mu &:&\mathcal{B}(\Sigma _{M})\rightarrow \lbrack 0,1]
\label{prob 2 prescript} \\
\mu (\sigma _{i}\cup \sigma _{j}) &=&\mu (\sigma _{i})+\mu (\sigma _{j}) 
\notag \\
\mu (\Sigma _{M}) &=&1  \notag
\end{eqnarray}

In order to calculate $p(x|\psi ^{s}),$ we need to evaluate the probability
measure over the set of states for the observer giving rise to the outcome $%
x $ when they interact with a state $\psi ^{s}$: 
\begin{eqnarray}
p(x|\psi ^{s}) &=&\mu (eig(x,\psi ^{s}))/\mu (\cup _{i=1}^{n}eig(x_{i},\psi
^{s}))  \label{probability} \\
&=&\mu (eig(x,\psi ^{s}))
\end{eqnarray}%
This last formula is fundamental to this paper. It says that for a repeated
experiment on a set of identical pure system states, the probability $%
p(x|\psi ^{s})$ is given as the ratio of observer states that, given $%
\psi ^{s},$ tell the outcome is $x,$ to the total number of observer states.

Note that the sets $eig(x_{i},\psi ^{s})$ are \emph{not} sets of
eigenvectors in the algebraic sense of the word\footnote{The sets (\ref{eigensets}) are called in eigensets in accordance with \cite{Aerts et al}.}. 
However, if it happens to be the case that, for a given $%
\psi ^{s}$ and for almost every $\psi \in $ $\Sigma _{M}$ , we have $\tau
(\psi ^{s},\psi )=x_{k}$ in the sense that 
\begin{equation}
\mu (eig(x_{k},\psi ^{s}))=\mu (\Sigma _{M})  \label{normalization 2}
\end{equation}%
then, for that particular $\psi ^{s},$ we have $p(x_{k}|\psi ^{s})=1.$ The
vector $\psi ^{s}$ thus defined, will coincide with a regular eigenvector if
the state space is a Hilbert space.

The relation between (\ref{prob 2 prescript}) and (\ref{prob prescript}) is
through the mapping $\tau $ and the measure $\mu .$ It is obvious that (\ref%
{prob 2 prescript}) is additive in $X$ too 
\begin{equation}
\mu (eig(x_{i},\psi ^{s})\cup eig(x_{j},\psi ^{s}))=\mu (eig(x_{i},\psi
^{s}))+\mu (eig(x_{j},\psi ^{s}))  \label{additivity}
\end{equation}%
because of (\ref{eigpart 1}). Hence, if the probabilities of (\ref{prob 2
prescript}) and (\ref{prob prescript}) coincide for every single outcome
(the singletons in $\mathcal{P}(X)$ ), they will coincide for all of $%
\mathcal{P}(X).$ In what follows we will therefore restrict our discussion
to the probability related to the occurrence of a \emph{single} outcome. In
conclusion, the success of the program to model the probabilities in quantum
mechanics as coming from a lack of knowledge about the precise state of the
observer stands or falls with the question of defining a natural mapping $%
\tau $ (which determines the outcome and hence $eig(x,\psi ^{s})$ ) such
that the measure $\mu $ of the eigenset $eig(x_{i},\psi ^{s})$ pertaining to
outcome $x_{i}$ is identical with the probability obtained by the Born rule (%
\ref{Born}).

\subsection{The Bayes-optimal observer}

We can see from (\ref{probability}) that the system state $\psi ^{s}$ can be
associated with a probability in a fairly trivial way: the probability of an
outcome $x$ when the system is in a pure state $\psi ^{s}$, is the
proportion of observer states that attribute outcome $x$ to that state. Even
for a repeated measurement on a set of identical pure states, fluctuations in the outcomes
can arise if there is a lack of knowledge concerning the precise state of the
observer. Suppose now the observer, considered as a system in its own right,
is in a state $\psi ^{m}$. Then in exactly the same way we can associate a
probability with that state too. The operational meaning of this association
is given either by a secondary observer observing an ensemble of observers
in the state $\psi ^{m}$, or by the observer consistently (mis)identifying
his own state $\psi ^{m}$ for a state of the system $\psi ^{s}$. We have
argued that every outcome will say something about the observer, (that is,
about $\psi ^{m}$), and something about the system (that is, about $\psi
^{s} $). The problem is that this information is mixed up in a single
outcome. Some outcomes will contain more information about the state of the
system, and some more about the state of the apparatus. Eventually,
we, as operators of our detection apparatus, will have to decide whether we
will retain a given outcome, or reject it. Such decisions are a vital part
of experimental science. For example, an outcome that is deemed too far off
the limit (so-called \emph{outliers}), is rejected and hence excluded in the
subsequent analysis. The rationale for this exclusion is that an outlier
does not contain information about the system we seek to investigate, but
rather that it represents a peculiarity of the measurement. In practice,
rejection or acceptance of an outcome does not depend on a rational
analysis, but on the common sense and expectations of the experimenter.
Suppose however, that the observer does have absolute knowledge about the
state of the system $\psi ^{s}$ and his own state $\psi ^{m},$ and
recognizes the fact that the outcome he delivers may eventually be rejected.
The observer considers this rejection to be based on the following binary
hypotheses:%
\begin{eqnarray}
H_{0} &:&\text{the outcome }x_{i}\text{ was inferred from }\psi ^{s}
\label{hypotheses} \\
H_{1} &:&\text{the outcome }x_{i}\text{ was inferred from }\psi ^{m}  \notag
\end{eqnarray}

In full, the hypotheses should actually read: \textquotedblleft The outcome $%
x_{i}$ yields as a consequence of the observer attributing the state $\psi
^{s}$ (or $\psi ^{m}$) to the system\textquotedblright . To combat
rejection, the observer chooses the outcome that maximizes the likelihood
that $H_{0}$ prevails, \emph{as if the outcome he delivers will eventually
be judged for acceptance or rejection by one with absolute knowledge about }$%
\psi ^{s}$\emph{\ and }$\psi ^{m}$. If, in an experiment, it is possible
with (non-vanishing probability) to get an outcome $x_{i}$ under either
hypothesis, then a factual occurrence of this outcome in an experiment
supports\emph{\ both} hypotheses simultaneously. What really matters in
deciding between $H_{0}$ and $H_{1}$ on the basis of a single outcome, is
not the probability of the correctness of each hypothesis itself, but rather
whether one hypothesis has become \emph{more likely} than the other as a
result of getting outcome $x_{i}$. From Bayesian decision theory \cite{Jaynes 2003},
we have that all the information in the data that is relevant for deciding
between $H_{0}$ and $H_{1},$ is contained in the so-called likelihood ratios
or, in the binary case, the\emph{\ odds }$\Lambda _{i}$:%
\begin{equation}
\Lambda _{i}=\frac{p(x_{i}|\psi ^{s})}{p(x_{i}|\psi ^{m})},i=1,\ldots ,n
\label{odds}
\end{equation}
In this last formula, the numerator and denominator are given by (\ref%
{probability}). We are now in position to state our proposed strategy for
the Bayes-optimal observer.

\begin{definition}[Bayes-optimal observer]
We call a system $M$ in a state $\psi ^{m}$ a \emph{Bayes-optimal observer}
iff, after an interaction with a system in a state $\psi ^{s},$ the state of 
$M$ will transform to a state that expresses the outcome $x_{i}$ that
corresponds to the maximal likelihood ratio $\Lambda _{i}$ (\ref{odds}).
\end{definition}

Picking the outcome $x_{i}$ from $X$ that maximizes the corresponding
likelihood ratio $\Lambda _{i},$ is simply optimizing the odds for $H_{0},$
given his information. This concludes our description of the observer. To
see what probability arises for a repeated experiment when an observer is
Bayes-optimal, we need a state space. We are especially interested in
complex Hilbert space, but we will first have a look at statistical mixtures.

\subsection{The Bayes-optimal observer for statistical mixtures}

If the conditional probabilities $p(x_{i}|\psi ^{s})$ are well-defined
(which we will just accept for now), we can make a summary of them in a
single vector $\mathbf{x}(\psi ^{s}):$ 
\begin{equation}
\mathbf{x}(\psi ^{s})=\sum_{i=1}^{n}p(x_{i}|\psi ^{s})x_{i}
\label{stat state}
\end{equation}%
First we define the convex closure of a number of elements $a_{1},\ldots
,a_{n}$ $\in A,:$ 
\begin{equation}
\lbrack a_{1},\ldots ,a_{n}]=\{a\in 
\mathbb{R}
^{n}:a=\sum \lambda _{i}a_{i},0\leq \lambda _{i}\in 
\mathbb{R}
,\sum \lambda _{i}=1\}  \label{convex closure}
\end{equation}%
If we write $[C],$ as we shortly will, we mean the convex closure of the
elements in $C.$ The standard $(n-1)$ simplex $\Delta _{n-1}$ generated by
the outcome set $X$ is:%
\begin{equation}
\Delta _{n-1}(X)=[x_{1},\ldots ,x_{n}]  \label{convex state space}
\end{equation}%
We see from (\ref{stat state}), (\ref{normalization 2}) that $\mathbf{x}%
(\psi ^{s})$ belongs to $\Delta _{n-1}(X).$ By identification of the axes of 
$
\mathbb{R}
^{n}$ with the members of $X,$ we have $\Delta _{n-1}(X)\subset 
\mathbb{R}^{n}(X),$ the free vector space generated by the outcome set $X$. Vectors
like $\mathbf{x}(\psi ^{s})$ are often called `statistical states' or
`mixtures' in the literature. Suppose now that all we can or care to know
about the system $S$ and the observer $M$, are the statistical states, i.e.
the probabilities related to the outcomes of a single experiment. Within
this constraint, the vector $\mathbf{x}(\psi ^{s})$ represents all there is
to know about $S$ and the state spaces $\Sigma _{S}$ and $\Sigma _{M}$
reduce to $\Delta _{n-1}(X):$ 
\begin{equation}
\Sigma _{S}=\Sigma _{M}=\Delta _{n-1}(X)  \label{state spaces}
\end{equation}%
Having identified $\psi ^{s}$ with $\mathbf{x}(\psi ^{s})$ in this
particular case, the conditional probability $p(x_{1}|\mathbf{x}(\psi ^{s}))$
denotes the probability that outcome $x_{1}$ occurs when our knowledge about
the system is encoded in the statistical state $\mathbf{x}(\psi ^{s}):$%
\begin{equation}
\mathbf{x}(\psi ^{s})=p(x_{1}|\mathbf{x}(\psi ^{s}))x_{1}+\ldots +p(x_{n}|%
\mathbf{x}(\psi ^{s}))x_{n}  \label{stat state 2}
\end{equation}%
In this section $\langle ,\rangle $ denotes the standard inner product in
Euclidean space, and with $\langle x_{i},x_{j}\rangle =\delta _{ij}$, we
have from this last equation 
\begin{equation}
p(x_{i}|\mathbf{x}(\psi ^{s}))=\langle \mathbf{x}(\psi ^{s}),x_{i}\rangle
\label{prob assignment}
\end{equation}%
For a statistical state, the magnitude of the $i^{th}$ coordinate equals the
probability of outcome $x_{i}.$ We have a state space (\ref{convex state
space}), and we have a rule to extract a probability from a state (\ref{prob
assignment}), so we can characterize the sets $eig(x_{k},\mathbf{x}(\psi
^{s})).$ Let $\mathbf{x}(\psi ^{s})$ and $\mathbf{x}(\psi ^{m})$ be
arbitrary states in $\Delta _{n-1}(X),$ written as: 
\begin{eqnarray}
\mathbf{x}(\psi ^{s}) &=&\sum_{i=1}^{n}t_{i}x_{i}  \label{vectors} \\
\mathbf{x}(\psi ^{m}) &=&\sum_{i=1}^{n}r_{i}x_{i}  \notag
\end{eqnarray}%
By the definition of Bayes-optimal observation, we have that the outcome $%
x_{k}$ is chosen, if for all $j\neq k,$ the corresponding likelihood ratio's
satisfy $\Lambda _{k}>\Lambda _{j}.$ By (\ref{odds}) and (\ref{stat state}), 
$x_{k}$ is chosen, iff for all $j=1,\ldots ,n\quad (j\neq k),$ we have: 
\begin{equation}
\frac{p(x_{k}|\mathbf{x}(\psi ^{s}))}{p(x_{k}|\mathbf{x}(\psi ^{m}))}>\frac{%
p(x_{j}|\mathbf{x}(\psi ^{s}))}{p(x_{j}|\mathbf{x}(\psi ^{m}))}
\label{BO char}
\end{equation}%
The regions $eig(x_{k},\mathbf{x}(\psi ^{s})),$ are found by substitution of
(\ref{vectors}) in (\ref{prob assignment}) and then into (\ref{BO char}).
With $j=1,\ldots ,n;\quad j\neq k$, we obtain: 
\begin{equation}
eig(x_{k},\mathbf{x}(\psi ^{s}))=\{\mathbf{x}(\psi ^{m})\in \Delta _{n-1}:%
\frac{t_{k}}{r_{k}}>\frac{t_{j}}{r_{j}}\}  \label{real eigensets}
\end{equation}%
According to (\ref{probability}), the probability of the outcome $x$ for the
repeated experiment on a set of identical system states, is the ratio of
observer states that tell the outcome is $x,$ to the total number of
observer states. Because the state space is Euclidean, it is natural to take
for $\mu $ the $(n-1)$-Lebesgue measure in $\Delta _{n-1},$ assumed to be
normalized: $\mu (\Delta _{n-1}(X))=1.$ The probability $p_{BO}(x_{k}|%
\mathbf{x}(\psi ^{s}))$ that the Bayes-optimal observer obtains the outcome $%
x_{k}$ is then given by%
\begin{equation}
p_{BO}(x_{k}|\mathbf{x}(\psi ^{s}))=\mu (eig(x_{k},\mathbf{x}(\psi ^{s})))
\label{PBO}
\end{equation}%
However, because of the way we defined the statistical state, the
probability is also given directly by components of the state. So the
question is whether the Bayes-optimal observer (\ref{PBO}) can recover that
probability, i.e. is it true that (\ref{PBO}) equals (\ref{prob assignment}%
): 
\begin{equation}
\mu (eig(x_{k},\mathbf{x}(\psi ^{s})))=\langle \mathbf{x}(\psi
^{s}),x_{i}\rangle  \label{criterion}
\end{equation}%
To see if this is the case, we first define the open convex closure of a
number of elements $x_{1},\ldots ,x_{n}$ $\in 
\mathbb{R}
^{n}$ as 
\begin{equation}
]x_{1},\ldots ,x_{n}[=\{x\in 
\mathbb{R}
^{n}:x=\sum \lambda _{i}x_{i},0<\lambda _{i}\in 
\mathbb{R}, \sum \lambda _{i}=1\}  \label{open closure}
\end{equation}%
We can now characterize $eig(x_{k},\mathbf{x}(\psi ^{s}))$ for the statistical
state as being `almost equal' to 
\begin{equation}
C_{k}^{s}=]x_{1},\ldots ,x_{k-1},\mathbf{x}(\psi ^{s}),x_{k+1},\ldots ,x_{n}[
\label{eigset}
\end{equation}

A graphical representation of the eigensets in the simplex state space can be found in Figure (1).

\begin{lemma}
Let $C_{k}^{s}$ be defined as in (\ref{eigset}), $[C_{k}^{s}]$ be the convex
closure of $C_{k}^{s},$ and $eig(x_{k},\mathbf{x}(\psi ^{s}))$ by (\ref{real
eigensets}), then: 
\begin{equation*}
C_{k}^{s}\subset \mathbf{\ }eig(x_{k},\mathbf{x}(\psi ^{s}))\subset \lbrack
C_{k}^{s}]
\end{equation*}
\end{lemma}

\begin{figure}[tbp]
\begin{center}
\epsfig{file=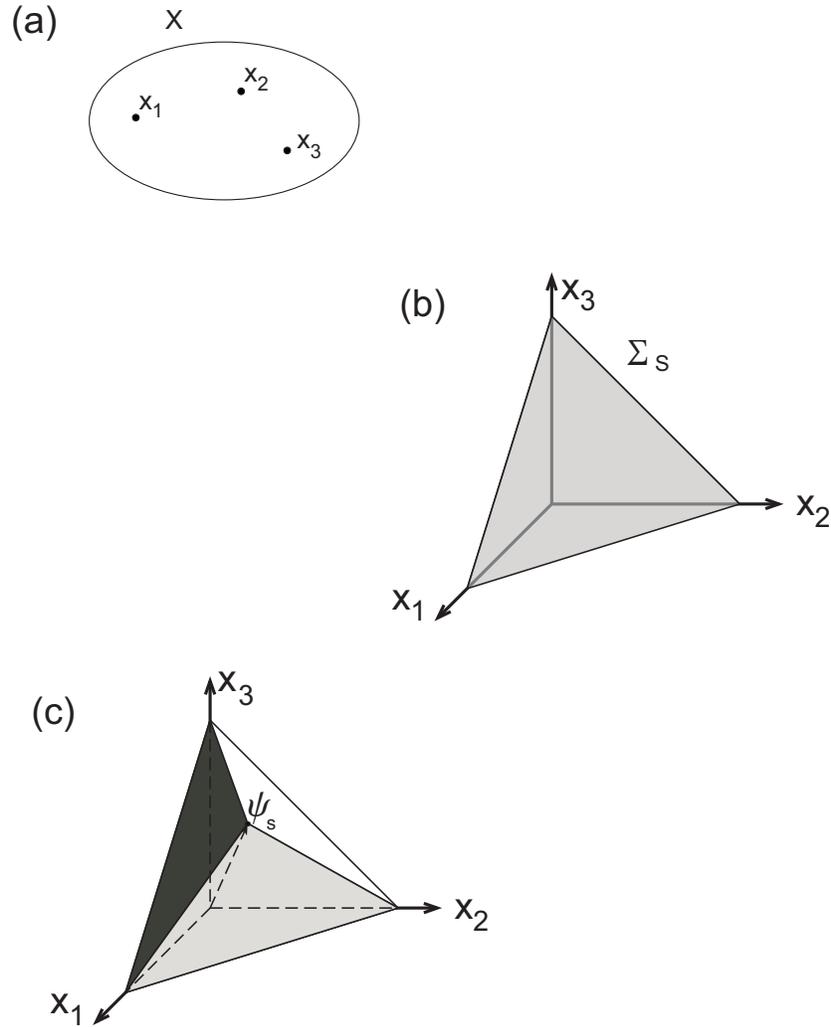, width=11cm}
\end{center}
\caption[Real state space]{Illustration of the scheme in the simplex state
space. We start with the discrete outcome set, depicted in figure (a). The
state space for an outcome set with three outcomes, is the standard
2-simplex in the free vector space generated by the outcome set over the
field of real numbers, as depicted in picture (b). In figure (c), we see the eigensets $C_{k}^{s}$, so we can see what outcome will be obtained from a
Bayes-optimal measurement. An apparatus state picked from the darkest region, $C_{2}^{s}$, 
will lead to the outcome $x_{2}$, in the lightest region, $C_{1}^{s}$,  to $x_{1}$, and the
intermediately shaded region, $C_{3}^{s}$, leads to the outcome $x_{3}$. The probability
is the Lebesgue measure over the depicted eigensets. I.e., the probability
of obtaining the outcome $x_{2}$, is the normalized area of the darkest
triangle.}
\label{Fig:real_hilbert}
\end{figure}

The proof of this lemma can be found in appendix A. To obtain the
probability (\ref{PBO}), we calculate the $\mu -$measure of $[C_{k}^{s}],$
which is simply the $(n-1)$-dimensional volume of the simplex $[C_{k}^{s}].$

\begin{lemma}
If $\mu $ is a (probability) measure such that $\mu (\Delta _{n-1}(X))=1,$
and $C_{k}^{s}$ is defined by the convex closure of (\ref{eigset}), then we
have $\mu ([C_{k}^{s}])=t_{k}$
\end{lemma}

One can calculate of the volume of a simplex straightforwardly by
determinant calculus, as was done in \cite{Aerts 1986}. For completeness, we
have included an alternative in the form of a simple geometric argument in
appendix B. We then easily obtain:

\begin{theorem}
$\mu (eig(x_{k},\psi ^{s}))=t_{k}$
\end{theorem}

\begin{proof}
By the first lemma, we have $C_{k}^{s}\ \subset \ eig(x_{k},\mathbf{x}(\psi
^{s}))\subset \lbrack C_{k}^{s}]$. Because $A\subset B\implies \mu (a)\leq
\mu (B)$ we have%
\begin{equation*}
\mu (C_{k}^{s})\mathbf{\ }\leq \mu (eig(x_{k},\mathbf{x}(\psi ^{s})))\leq
\mu ([C_{k}^{s}])
\end{equation*}%
By the second lemma we have $\mu ([C_{k}^{s}])=t_{k}$. To calculate $\mu
(C_{k}^{s}),$ we note that $\mu (C_{k}^{s})=\mu ([C_{k}^{s}])-\mu
([C_{k}^{s}]\cap C_{k}^{s}).$ Because $[C_{k}^{s}]\cap C_{k}^{s}$ is the
collection of faces of $C_{k}^{s},$ a set of finite cardinality whose
members have an affine dimension maximally equal to $n-2,$ it is $\mu -$%
negligible, hence we also have $\mu (C_{k}^{s})=t_{k},$ establishing the
result.
\end{proof}

We see that indeed the Bayes-optimal observer recovers the probability that
was encoded in the statistical state: 
\begin{equation*}
p_{BO}(x_{k}|\mathbf{x}(\psi ^{s}))=t_{k}=p(x_{k}|\mathbf{x}(\psi ^{s}))
\end{equation*}

In this way the observer succeeds in obtaining a quantity that, in the limit
of infinite measurements, depends only on the state of the system under
investigation, and not on his own state. The results we have obtained for
the simplex state space are identical to those in \cite{Aerts 1986}, where
the scheme was proposed under the name \textquotedblleft hidden
measurements\textquotedblright\ to indicate the origin of the lack of
knowledge. In \cite{Aerts 1986} the eigensets are postulated \emph{ad hoc},
whereas we have derived their simplicial shape from the principle of
Bayes-optimal observation. We will use this principle in the next section to
extend the results of \cite{Aerts 1986} to systems with a complex state
space.

\begin{figure}[tbp]
\begin{center}
\epsfig{file=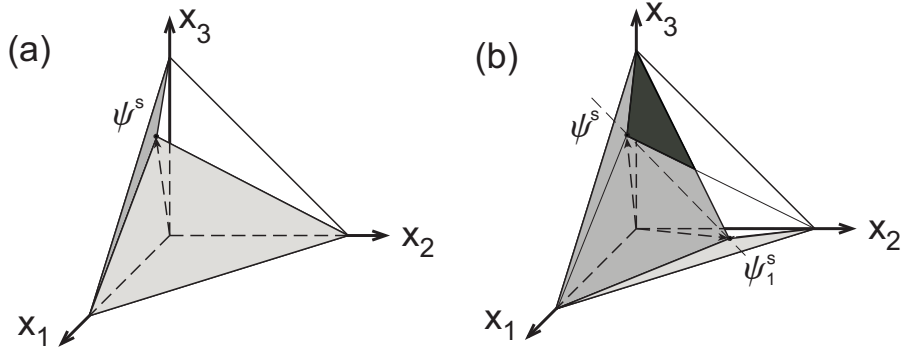, width=12cm}
\end{center}
\caption[Contextuality of the outcome assignment.]{A graphical exposition of
the contextuality of the outcome assignment. (a) If we pick an observer
state from the open white triangle $] \protect\psi^{s},x_{2},x_{3}[$, then a
measurement of the state $\protect\psi _{s}$ will yield outcome $x_{1}$. (b)
If we interchange the second and third component of $\protect\psi ^{s}$, we
obtain $\protect\psi _{1}^{s}.$ The probability of obtaining outcome $x_{1}$
is the same as in picture (a), because the two triangles have the same area.
However, an observer state choosen from the black shaded region would yield
outcome $x_{1}$ in picture (a), whereas it would yield $x_{2}$ in picture
(b). Note that we did not change the $x_{1}$ component in the state $\protect%
\psi^{s}$ to obtain $\protect\psi _{1}^{s}.$ }
\label{picture contextual}
\end{figure}

Before we do so, two remarks are in order. First, we did not specify whether
the state $\mathbf{x}(\psi ^{s})$ is the result of mixing `pure' components
with appropriate weights, as indicated by the components of the state, or
whether it represents a statistical tendency, somewhat like a propensity, of an ensemble of identical
`pure' states to reveal itself in the different outcomes. 
That is, if all we
are allowed to do is perform a single experiment on each member of the
ensemble, then from the resulting statistics of a single observable, we
cannot distinguish between these two situations. In other words, if we have
an urn filled with coins and we are allowed to inspect the coin only after a
single throw of the coin, for every coin in the urn, then we cannot know
whether it is a tendency of the coin to show heads with probability 1/2, or
whether half of the coins have both sides heads and half of them have both
sides tails (or indeed a mixture of these two situations). Secondly, it is
interesting that, even for the conceptually simple statistical mixtures, the
outcome assignment given by the Bayes-optimal observer is \emph{contextual }%
in the following sense: given a state for the observer and system that lead
to the outcome $x_{l},$ then the mere interchanging of the coefficients $%
t_{j}$ and $t_{k}$ (equal to the probability for the outcomes $x_{j}$ and $%
x_{k}$) can easily result in a different outcome than $x_{l},$ even if \emph{%
neither} $x_{j},$ \emph{nor} $x_{k}$ is equal to $x_{l}$! This can readily
be verified in Figure (\ref{picture contextual}). However, the probability $p(x_{i}|%
\mathbf{x}(\psi ^{s}))$ of the outcome is a function of $t_{i}$ only, hence
the probability itself is non-contextual.
Conversely, given a state of an observer $\psi ^{m}$ and a system state $%
\psi ^{s}$ that interact to yield the outcome $x_{k},$ it is often possible
to change the outcome of the Bayes-optimal observer to a different outcome
by interchanging suitable coefficients of the observer, \emph{leaving }$%
r_{k} $\emph{\ untouched}. This means that changing \emph{only} the observer's
preferences over the outcomes $x_{j}$ and $x_{l},$ may let the Bayes-optimal
observer decide another outcome than $x_{k}$ is more optimal, even if $j,l$ 
and $k$ are all different! This contextual aspect of the outcome assignment can here be
understood as a result of the inescapable bias introduced by the state of the observer in producing a single outcome, for the coefficients of
his state represent his tendencies for each outcome\footnote{This tendency could be revealed if we fix the state of the observer and observe (by means of a second observer)the 
relative frequencies for the outcomes he produces when he measures members of an ensemble of randomly choosen states.}. Perhaps somewhat paradoxically, it is
precisely through the averaging procedure over all the different
possibilities for this bias, that a non-contextual probability emerges. From
Figure (\ref{picture contextual}) we see that the contextuality of the outcome
assignment depends on the classical entropy of the state. According to a
well-known theorem due to Shannon, the higher the entropy of the state $\psi
^{s},$ the closer the coefficients of $\psi ^{s}$ in (\ref{stat state}) are
to $1/n$ and the closer this state will reside near the centre of the
simplex, effectively limiting the possibilities for producing a contextual
outcome change by interchanging coefficients.

\subsection{Bayes-optimal observation in complex Hilbert space}

Complex Hilbert spaces are of considerable interest as they arise naturally
in many prominent scientific areas including quantum theory, signal analysis
(both in time-frequency and in wavelet analysis), electromagnetism and
electronic networks\footnote{%
Interestingly, the name probability \emph{amplitude}, and indeed the Born
interpretation of the wave vector in quantum mechanics, were conceived by
Born in analogy with electromagnetic waves. Here the norm is not unity, but
equal to the energy in the wave, and probability conservation is replaced by
conservation of energy.}, and the more recently founded shape theory \cite%
{Kendall 1999}. The natural setting for the discrete state space in these
examples, is the space of square summable functions on a Hilbert space $%
\mathcal{H}_{n}(%
\mathbb{C}
)$ over the field of complex numbers. A general state of the system $\psi
^{s}\in \Sigma _{S}=\mathcal{H}_{n}(%
\mathbb{C}
)$ can then be written as:%
\begin{equation}
\psi ^{s}=\sum_{i=1}^{n}q_{i}x_{i}  \label{complex state}
\end{equation}where $q_{i}\in  \mathbb{C}$ and $|\psi ^{s}|=1.$ In this case the outcome set $X$ consists of an
orthonormal frame of complex vectors $\{x_{i}\}.$ An observer (or a
detector, which is quite the same for our purposes) usually has a very large
number of internal degrees of freedom. Accordingly it lives in a Hilbert
space of appropriately high dimensionality. However, by the Schmidt
bi-orthogonal decomposition theorem, we know we can model every possible
interaction between two systems, one living in a Hilbert space of dimension $%
n$ and one in a Hilbert space of dimension $m$ with $m>n,$ by an interaction
of two systems, each one living in a Hilbert space of dimension $n.$ With
this in mind, we model the set of states of the observer as unit vectors in $%
\mathcal{H}_{n}$: 
\begin{equation*}
\Sigma _{M}=\{\psi \in \mathcal{H}_{n}(\mathbb{C}) :|\psi |=1 \}
\end{equation*}

\begin{figure}[tbp]
\begin{center}
\epsfig{file=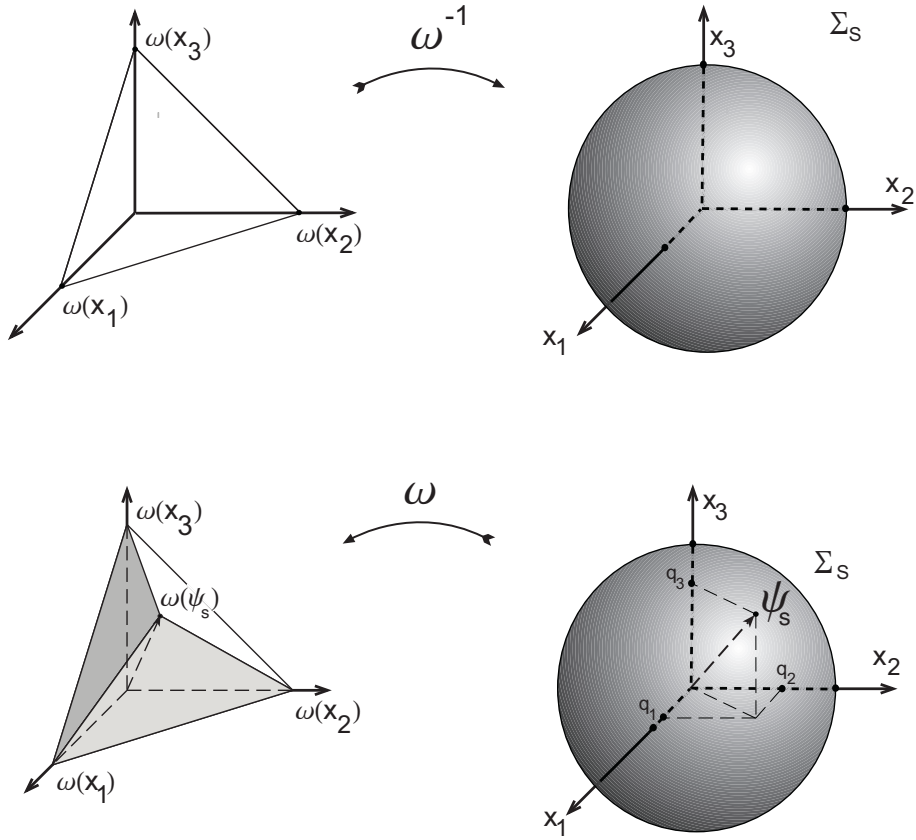,  width=\linewidth}
\end{center}
\caption[Complex state space.]{The action of the mapping $\protect\omega $
sends elements of the unit sphere to the standard simplex (upper figure).
The probability for the occurrence of outcome $x_{k}$ is the measure of the
eigenset corresponding to outcome $x_{k}$ and is calculated in the simplex
using the measure preserving mapping $\protect\omega $. The eigensets are
depicted in the lower figure for the simplex; it is not possible to show
graphically what these sets look like in the complex unit sphere.}
\label{picture complex}
\end{figure}

The reader should take note of the fact that, every time we speak about
\textquotedblleft the state of the observer\textquotedblright , we mean the
state in the subspace indicated by the Schmidt bi-orthogonal decomposition
theorem. The state of the observer, to us, always means only that part of
the state that is of relevance to the production of the outcome. This is
especially relevant for the interpretation of sentences such as
\textquotedblleft uniform distribution of initial observer
states\textquotedblright , which taken too literally, would indicate the
observer is perhaps doing something completely different than observing. The
state of an observer with respect to an experiment with outcome set $X$ can
be written as ($r_{i}\in 
\mathbb{C}
$) 
\begin{equation}
\psi ^{m}=\sum_{i=1}^{n}r_{i}x_{i}  \label{complex observer state}
\end{equation}%
Because the coefficients now assume complex values, they cannot be
interpreted as probabilities because we do not have a total order relation
in the field of complex numbers. This difference also affects the deeper,
deterministic level of the description in a profound way. Let us explain why
this is the case. For the statistical states of the former section, each
eigenset is a subsimplex of the state space. A simplex is a (very) special
case of a convex set. Because the eigensets share at most a lower
dimensional face, any two different eigensets (for a fixed system state) can
be separated\footnote{%
If $C_{1}$ and $C_{2}$ are two sets in $\mathbb{R}^{n},$ then a hyperplane $%
H $ is said to \emph{separate }$C_{1}$ and $C_{2}$ iff \emph{\ }$C_{1}$ is
contained in one of the closed halfspaces associated with $H$ and $C_{2}$
lies in the opposite closed half-space.\ Two convex sets in $\mathbb{R}^{n}$
that share at most an affine set of dimension $n-1,$ can be separated by a
hyperplane.} by a single hyperplane. But in a complex space a hyperplane
does not separate that space in two half-spaces. To apply the criterion of
Bayes-optimality, one needs to decomplexify the space to restore the order
relation, but this can be done in a variety of ways. On the other hand, this
plurality of decomplexifications need not bother us too much. Just as in the
case of the statistical states of the former section, the observer can check
the statistical validity of his outcome assignment by verifying that the
probability (in the sense of a relative frequency) that results from
repeated application of his outcome assignment, equals the \emph{assumed}
probability. In the same way, we can simply postulate, or even guess, a
specific form of the probability assignment and justify it \emph{a posteriori%
}: If the relative frequency of an outcome (as a result of the observers'
outcome assignment, based on the Bayes-optimal condition), converges to a
limit that yields (a monotone function of) the very probability assignment
he used to obtain those outcomes, the Bayes-optimal observer knows he was
Bayes-optimal. Let us attempt a minimal generalization of the real case (\ref%
{real eigensets}), with $\psi ^{s}$ and $\psi ^{m}$ defined as in (\ref%
{complex state}), (\ref{complex observer state}) and $j=1,\ldots ,n;\quad
j\neq k$: 
\begin{equation}
eig^{\mathbb{C}}(x_{k},\psi ^{s})=\{\psi ^{m}\in \Sigma _{M}:\frac{|q_{k}|}{|r_{k}|}>\frac{%
|q_{j}|}{|r_{j}|}\}  \label{eig complex}
\end{equation}%
The only difference with (\ref{real eigensets}), is that we take the \emph{%
modulus} of the coefficients and that the set contains complex vectors,
which is why we have given the eigenset the superscript $\mathbb{C}$. 
To check the consistency of our Bayes-optimal observer in the complex
state space, we evaluate the Lebesgue measure $\nu (eig^{\mathbb{C}}(x_{k},\psi ^{s})).$ 
Therefore we regard the measure $\nu $ in 
$\mathbb{C}^{n}$ as the Lebesgue measure $\mu $ over $\mathbb{R}^{2n}.$ 
The calculation of the measure by direct integration can be avoided
by use of a mapping $\omega $ that preserves measures. A measurable mapping $%
\omega $ between measure spaces $\left( \Sigma ,\mathcal{A},\mu \right) $
and $(\Sigma ,\mathcal{B},\nu )$ is called a measure-preserving mapping if,
for every $B\in \mathcal{B}$, we have $\mu (\omega ^{-1}(B))=\nu (B)$. In
appendix C we demonstrate that the component-wise (or Haddamard) product of
a complex vector with its complex conjugate, that sends elements of the
complex unit-sphere $S_{n}=\{z\in \mathbb{C}^{n}:\sum_{i=1}^{n}z_{i}z_{i}^{%
\ast }=1\}$ onto the $(n-1)$ -simplex $\Delta _{n-1}=\{x\in \mathbb{R}%
_{+}^{n}:\sum_{i=1}^{n}x_{i}=1\}$ is indeed measure preserving in this
sense. We have given a graphic representation of the action of $\omega $ in
Figure (\ref{picture complex}). We are now in a position to prove our main
result.

\begin{theorem}
\begin{equation*}
p(x_{k}|\psi ^{s})=|\langle x_{k},\psi ^{s}\rangle |^{2}
\end{equation*}
\end{theorem}

\begin{proof}
With $eig^{\mathbb{C}}(x_{k},\psi ^{s})$ defined by (\ref{eig complex}), and 
\begin{equation*}
C_{k}^{s}=]\omega (x_{1}),\ldots ,\omega (x_{k-1}),\omega (\psi ^{s}),\omega
(x_{k+1}),\ldots ,\omega (x_{n})[,
\end{equation*}%
it is straightforward to show that (for more details, see \cite{Aerts Born})
we have:%
\begin{equation*}
C_{k}^{s}\subset \omega (eig^{\mathbb{C}}(x_{k},\psi ^{s}))\subset \lbrack C_{k}^{s}]
\end{equation*}%
Let $\tilde{\mu}$ and $\tilde{\nu}$ stand for the normalized versions of the
measures $\mu $ and $\nu $ in the proof in appendix C, so that their
constant of proportionality equals one: $\tilde{\nu}(\omega ^{-1}(A))=\tilde{%
\mu}(A).$ By definition $p(x_{k}|\psi ^{s})=\tilde{\nu}(eig^{\mathbb{C}}(x_{k},\psi ^{s}))$, and by the previous lemma, we have%
\begin{eqnarray*}
\tilde{\nu}(eig^{\mathbb{C}}(x_{k},\psi ^{s})) &=&\tilde{\nu}(\omega ^{-1}(C_{k}^{s})) \\
&=&\tilde{\mu}(C_{k}^{s})
\end{eqnarray*}%
The normalized measure $\tilde{\mu}(C_{k}^{s})$ of the real simplex $%
C_{k}^{s}$ was calculated in the real state space. A completely equivalent
calculation gives us%
\begin{eqnarray*}
\tilde{\mu}(C_{k}^{s}) &=&\langle \omega (x_{k}),\omega (\psi ^{s})\rangle
=|q_{k}|^{2} \\
&=&|\langle x_{k},\psi ^{s}\rangle |^{2}
\end{eqnarray*}
\end{proof}

We see that indeed the Bayes-optimal observer recovers the Born rule as a
result of his attempt to maximize the odds with respect to the outcome that
pertains to the system. To be precise, we did not maximize the odds, because
substitution of the Born rule for the probability in (\ref{odds}) gives:%
\begin{equation}
\Lambda _{k}=\frac{|\langle x_{k},\psi ^{s}\rangle |^{2}}{|\langle
x_{k},\psi ^{m}\rangle |^{2}}=\frac{|q_{k}|^{2}}{|r_{k}|^{2}}
\label{QM odds}
\end{equation}%
Whereas our observer, by (\ref{eig complex}), calculated the ratio's: 
\begin{equation}
\tilde{\Lambda}_{k}=\frac{|q_{k}|}{|r_{k}|}  \label{pseudo odds}
\end{equation}

where the tilde denotes the fact that, strictly speaking, this is not a
likelihood, because $|q_{k}|$ and $|r_{k}|$ aren't probabilities (they are
square roots of probabilities). Yet, it is obvious that the value of $k$ for
which (\ref{QM odds}) and (\ref{pseudo odds}) are maximal, is the same
because one is the square of the other, which is clearly a monotone
function. As a consequence, it does not matter if the Bayes-optimal observer
works with (\ref{QM odds}) or with (\ref{pseudo odds}): repeated application
of either strategy on the same pure state will make the relative frequency
converge to the Born rule in exactly the same way in both cases.

\section{Consequences of Bayes-optimal observation}

\subsection{Decision invariance and unitarity}

The outcome chosen by a Bayes-optimal observers, is the one that maximizes
the corresponding likelihood ratio $\Lambda _{i}.$ Any monotonously
increasing function of the likelihood ratio's preserves their relative
order, and hence their maximum. By (\ref{real eigensets}) and (\ref{eig
complex}), this carries over to the coefficients of the state vectors in
both the real and the complex state space. The same is true for
multiplication by a phase factor, which is cancelled by taking the moduli in
(\ref{eig complex}). As a result, the state space is not only a vector
space, it is a \emph{projective vector space}: if the vectors in the state
space are multiplied $z\in 
\mathbb{C},$ $0<|z|<\infty $, this does not change the result of the decision
procedure adopted by the Bayes-optimal observer. There is another
interesting class of transformations that leaves the Bayes-optimal decision
unaltered. For any $\psi ^{s}$, the probability of $x_{k}$ is defined as: 
\begin{equation*}
p(x_{k}|\psi ^{s})=\mu (eig^{%
\mathbb{C}}(x_{k},\psi ^{s}))
\end{equation*}%
Because $\omega (eig^{%
\mathbb{C}}(x_{k},\psi ^{s}))\subset \lbrack C_{k}^{s}],$ $\omega $ continuous, and
because the elements of $[C_{k}^{s}]$ have finite norm, the norm of the
vectors in $eig^{%
\mathbb{C}}(x_{k},\psi ^{s})$ is finite too. We can then apply a linear transformation
to the base vectors of the state space:%
\begin{eqnarray}
T &:&\Sigma _{S}\rightarrow \Sigma _{S}  \label{linear transformation} \\
T(x_{j}) &=&\sum_{i}^{n}\sigma _{ij}x_{j}  \notag
\end{eqnarray}

The eigenset $eig^{%
\mathbb{C}}(x_{k},\psi ^{s})$ will accordingly be transformed by applying $T$ to $%
x_{k} $ and $\psi ^{s}.$ By Lebesgue measure theory, the volume of the
transformed set is proportional to the volume of the original set, the
constant of proportionality being the determinant of the transformation: 
\begin{equation*}
\mu (T(eig^{\mathbb{C}}(x_{k},\psi ^{s})))=|\det (T)|\mu (eig^{%
\mathbb{C}}(x_{k},\psi ^{s}))
\end{equation*}%
for all $eig^{%
\mathbb{C}}(x_{k},\psi ^{s})\in \mathcal{B}(\Sigma ).$ This is a classic result%
\footnote{%
As before, we regard the complex $n-$space as a real $2n-$space, for which
the theorem is applicable.}, and we refer the interested reader to (\cite%
{Rudin 1987}, p54) for a proof. Note that this would typically be untrue for
a nonlinear transformation. As a result, all transformations with $|\det
(T)|=1$ leave the probabilities invariant, which means we have invariance
under\emph{\ unitary} transformations. Intuitively this is obvious: if the
probabilities have their origin in a measure on state space, then scaling,
phase shifting, forming the mirror image, or `rotating' the entire state space, does not alter the
relative proportions of the eigensets, hence the invariance. Of course, it
is easy to derive from the Born rule that the probabilities are invariant
under unitary transformations, because the Born rule is the square modulus
of an inner product and a unitary transformation can be defined as a linear
operator that leaves the inner product invariant. Our invariance principle
tells us the same story at a deeper level, for not only the probabilities
are invariant under unitary transformation, but also each obtained outcome
will be the same whether or not we unitarily transform the eigensets.

\subsection{The elusive quantum to classical transition}

Suppose we have a particular statistical mixture 
\begin{equation}
\varphi =\xi \psi _{1}+(1-\xi )\psi _{2}  \label{mixture}
\end{equation}%
of two (pure) states $\psi _{1}$ and $\psi _{2}$ with $\xi \in ]0,1[$. Suppose furthermore that 
\begin{eqnarray*}
p(x_{i}|\psi _{1}) &=&q_{1} \\
p(x_{i}|\psi _{2}) &=&q_{2}
\end{eqnarray*}%
Then an observing system is said to satisfy the\emph{\ linear mixture
property} iff%
\begin{equation}
p(x_{i}|\varphi )=\xi q_{1}+(1-\xi )q_{2}  \label{mixture property}
\end{equation}

In words: the probability of a mixture equals the mixture of the
probabilities. Does the Bayes-optimal observer satisfy the linear mixture
property? Well, $\varphi $ is a statistical mixture, as defined in the
section on Bayes-optimal observation of statistical mixtures, and each of
the constituents in the mixture is a pure state, as defined in the section
Bayes-optimal observation in Hilbert space. So clearly, our Bayes-optimal
observer satisfies the linear mixture property. In essence, this stems from
his initial states being uniformly random (almost everywhere). Indeed,
suppose the distribution of the initial states of the observer is \emph{not}
uniform a.e.. Then one can always find a convex region $S$ in state space
with surface measure $A$, for which the density of observer states is not
equal to $1/A.$ Without giving a formal proof, one can see that, it is
always possible to find two states $\psi _{1}$, $\psi _{2}\notin S$ and a
real number $\xi \in ]0,1[$, such that $\xi \psi _{1}+(1-\xi )\psi _{2}\in $ 
$S$ and for which the linear mixture property will be violated.

The linear mixture property is essential to experimental observation: no
experimenter would put his faith in the hands of a detection apparatus that
manifestly fails this most basic requirement. From this perspective, the
difficulty of finding an intermediate region between the classical and the
quantum, originates from the lack of a principle that determines \emph{how}
the observer should behave in order to objectively observe the intermediate
region in absence of the linear mixture property. As an example, suppose we
want to determine the length of a linearly extend system. In a classical
setting, we are in principle free to choose the number of outcomes, and we
are allowed to make many observations before we settle on the result of a
single measurement. For example, we can align the zero of the measuring rod
with one end point of the system and read the outcome at the other end point
as many times as we want to. If we are not satisfied with the precision that
the measuring rod affords, we can pick a better one, or improve it by adding
a nonius (or vernier) system to it. As long as we are able to do this, we
are still in a classical regime of observation. In the classical regime of
observation, the distribution of observer states will be highly non-uniform.
Ideally, of all possible measurements, the only uncertainty we have about
the state of the observer that is assumed to be of relevance to the
measurement outcome, is an uncertainty of the order of the smallest number
the measuring rod can represent. To decrease the uncertainty about the
result, even beyond the precision offered by the smallest number the rod can
represent, it is common scientific practice to perform the measurement many
times. Assuming identical, independent observations, one can apply standard
error theory. In the beginning of the eighties, Wootters has shown (\cite%
{Wootters 1980}, \cite{Wootters 1981}), using standard error theory, that
the distance (angle) between two states on the unit sphere in (real) Hilbert
space, is proportional to the number of maximally discriminating
observations along the geodesic between those two points. This beautiful
result gains in richness when considered from the point of view that the
probabilities arise in a Bayes-optimal way. In our search for ever more
precise measurements or measurements on ever smaller constituents of nature,
we eventually reach a region where we cannot repeat measurements without
absorbing the system or altering its state. We may not even be able to
choose freely the set of outcomes for a particular measurement, as is the
case in the quantum regime. It is then no longer possible to directly obtain
the ``true'' value of a physical quantity, because the eigenstate of the
observing system may not (and in general will not) coincide with the state
of the system under investigation. We cannot attempt the same measurement
(or one with altered eigenstates) on the same system, because the state of
the system has been altered, or even destroyed. In view of this
impossibility, we are led to statistical observation on ensembles. We have
shown it is possible to recover an objective probability \emph{if} the
distribution of observer states is uniform. We see that the best possible
observation scheme in the classical regime entails a \emph{minimal}
uncertainty (i.e. about the interpretation of the last digit only) in the
state of the observer, and in the quantum regime a \emph{maximal}
uncertainty (any outcome is in principle possible) about the state of the
observer. The consequence of such an interpretation is, that we will
only be able to identify intermediate regions when we allow for a more
complete description of the observing system. In essence, we need to
describe how to go from this minimal to this maximal uncertainty state.
There are good reasons for cautiously entering this intermediate region.
Some of the beautiful properties of the classical and the quantum regime
will not hold. For example, the linear mixture property cannot be
universally satisfied. Moreover, we will obtain probability distributions
that depend not only on the system, but at least partially on the dynamics
of the observing system. It is possible to construct explicit models that
show \cite{Aerts et al} one can identify an intermediate region where the
probabilities satisfy neither the classical statistical Bonferroni
inequalities\footnote{%
The Bonferroni inqualities indicate when a set of (joint) probabilities can
be derived from a Kolmogorovian probability model. The best known example of
a Bonferroni inequality in the foundations of quantum mechanics, is the Bell inequality.}
indicating the absence of a straightforward Kolmogorovian model, nor the
Accardi-Fedullo inequalities \cite{Accardi} that constrain the set of probabilities that
are derivable from a Hilbert space model. This opens up a whole new area of
investigation, but only if we are willing to take the bold step of
abandoning the full generality of the linear mixture property.

\subsection{Is the Bayes-optimal observer objective?}

The purpose of objective observation is to obtain a probability for the
outcome that depends only on the system under study. How fast the sequence
of outcomes converges to this probability, depends on how well the observer
manages to distinguish his state from the state of the system under study.
This aspect was neglected in the previous discussion. If we apply the Born
rule to calculate the quantities $p(x_{i}|H_{0})$ and $p(x_{i}|H_{1}),$ we
imply that $\sum_{i}p(x_{i}|H_{0})=\sum_{i}p(x_{i}|H_{1})=1.$ However, if
the choice between $H_{0}$ and $H_{1}$ is indeed a binary decision problem,
we should have:%
\begin{eqnarray}
\sum_{i}p(x_{i}|H_{0}) &=&\alpha  \label{def alpha} \\
\sum_{i}p(x_{i}|H_{1}) &=&1-\alpha  \notag
\end{eqnarray}

The reason why this is not contradictory, is
because the observer chooses his outcome, \emph{as if} the outcome will be
judged afterwards as a binary decision problem. The observer himself has a priori no
clue what the value of $\alpha $ might be. But even if he would estimate the value of $\alpha $ 
after repeated measurements, then still this knowledge cannot not help him to give a more optimal outcome. 
To the Bayes-optimal observer, knowledge of  $\alpha$  would merely have the
effect of scaling the odds in (\ref{odds}) by $\frac{1-\alpha }{\alpha }.$
The choice of the outcome for the Bayes-optimal observer is based on the
maximal likelihood and a monotone function of the likelihoods will not
change the maximum. Thus we see that the specific value of $\alpha $ has no
influence on the actual choice. If $\mu $ is truly uniform, then, the
resulting relative frequency will converge to the Bayes-optimal probability
that only depends on the state of the system, whatever value $\alpha $
happens to have in practice. However, a small value of $\alpha $ implies
that for each outcome, the probability that the outcome depends on the state of the system, is
small. So the expected increase in information about the system as a result
of obtaining that outcome, is small too. Evidently this will extend the number
of measurements needed to acquire information about the system. 
We see that $\alpha $ is a crude statistical measure for the
objectivity of the observer. It represents his ability to separate interior
from exterior. It turns out we can always pick an outcome that supports $%
H_{0}$ more than it supports $H_{1}$ iff $\alpha >1/2$. To see this, we
proceed ad absurdum. If no outcome supports $H_{0}$ more than it supports $%
H_{1},$ then for all $x_{j},$ 
\begin{equation*}
\frac{p(x_{j}|H_{0})}{p(x_{j}|H_{1})}\leq 1
\end{equation*}%
But then\footnote{%
This specific condition is known in the literature as majorization. It plays
an important role in the investigation of bipartite state conversions by
local operations and classical communications (LOCC). This may seem relevant
in connection to our problem, as the basic scheme we present can be
described as a bipartite state conversion problem. However, we cannot use
the many interesting results in the literature on bipartite state conversion
because LOCC's in this particular problem are operationally defined by means
of local unitary transformations and a local measurement, and it is the
local measurement that we seek to understand!} we have:%
\begin{equation}
\frac{\sum_{j}^{n}p(x_{j}|H_{0})}{\sum_{j}^{n}p(x_{j}|H_{1})}\leq 1,
\label{majorization}
\end{equation}%
which implies $\alpha \leq 1-\alpha .$ We obtain the contradiction iff $%
\alpha >1/2.$ In words: if we can do only slightly better than completely
arbitrary in letting the outcome probability depend on the system, we can
guarantee the existence of an outcome that maximizes the odds and is greater
than unity. In fact, for any value of $\alpha $ we can find an (almost
always unique) outcome that maximizes the odds, but when $\alpha >1/2,$ the
maximal likelihood ratio enjoys the property of being greater than one.

\subsection{The Bayes-optimal observer as a paradigm for observation}

The proposed principle of observation is based on a Bayesian treatment of a
binary decision problem, but is not used in its usual decision-theoretic
form. In decision theory we seek to establish which of the hypotheses enjoys
the strongest support in evidence of the data. In our case, there is no data
to feed the likelihood with, because we produce the data by means of the
odds. The way we employ the principle is like an inverse decision problem,
as if anticipating that the result will be judged afterwards by a decision
procedure performed by one with absolute knowledge of the system and
observer states prior to the measurement. The possibility of applying
Bayesian decision theory in quantum mechanics came to me through the
realization that the criterion established by Aerts D. at the end of \cite%
{Aerts 1986} to characterize the so-called hidden measurements, is a
monotone function of the Bayesian odds and hence leads to the same choice
for the outcome. In this sense, this paper can be seen as providing a
Bayesian foundation for the structure of the hidden measurements as given in, for example
 \cite{Aerts 1986} and \cite{Aerts 1987}, and extending the results to the complex Hilbert space.

More recently it has come to my attention that a somewhat similar paradigm
(without reference to quantum mechanics) is proposed in several papers that
deal with visual perception by humans. The idea that the visual system is
rooted in inference, can be traced back to the work of Helmholtz \cite%
{Helmholtz}, who proposed the notion of \emph{unconscious inference}. It was
only in the last decade that it was accepted and translated into a
mathematical framework, not in the least because computer scientists who
want to model the human vision system are faced with the apparent complexity
that underlies human perception. The Bayesian framework provides the tools
necessary to understand and explain a wide variety of sometimes baffling
visual illusions that occur in human perception \cite{Geisler Kersten}. In
retrospect, we have borrowed the term `Bayes-optimal' from this literature,
because the term so neatly describes the principle and it did not seem
appropriate to introduce a new term. There are however some differences in
the application of the principle with respect to our proposal. In the
literature on visual perception, the prior distributions are derived from
real world statistics. Of course, this begs the question how these prior
distributions were obtained in the first place. There are two basic
possibilities to obtain a prior: either a prior distribution is based on
some theoretical assumption, or it is established by looking at the relative
frequency of actual recordings. The first option is the one we pursued in
this article, where we assumed a uniform distribution of observer states%
\footnote{%
The absence of a more informative prior distribution effectively reduces the
criterion of Bayes-optimality to a Neyman-Pearson maximum-likelihood
criterion.}. In the second case, which is the one adopted in the literature
on perception, one has the advantage of being able to explain a wide variety
of visual effects in human perception, and how the priors can be adapted
through the use of Bayesian updating, but we cannot explain observation
itself. The relative frequency needed to obtain the prior, is rooted in the
observation of data, which requires another prior and so on ad infinitum.
One can break from this loop by reconsideration of what a state is. In the
literature on perception states are considered only as (real) statistical
mixtures, severely limiting both the applicability and the philosophical
scope of the paradigm. The state, as we have defined it here, can be a
complex vector, not obtainable as a mixture in principle, and yet give rise
to probabilities if we attempt to observe it as good as possible. So the
state is simultaneously a description of the `mode of being' (the pure state
that physically interacts), \emph{and} a `catalogue of information' (the
probabilities the Bayes-optimal observer obtains).

The possibility that the same principle governs human perception and quantum
mechanical observation, strengthens the Bayes-optimal paradigm. Measurement
apparata and human perception can be rooted in the same principle: the
attempt to relate the outcome to the object under investigation as
unambiguously as possible by choosing the outcome that has the largest odds (%
\ref{odds}). By repeating the observation many times, each time randomizing
the internal state of the sensor, we obtain an invariant of the observation
that pertains solely to the system.

Another interesting link with the existing literature was pointed out to me
by Thomas Durt \cite{Durt}. The regions of the Bohm-Bub model \cite{Bohm Bub}
coincide with our definition of the eigensets in the complex case (\ref{eig
complex}). Moreover, Bohm and Bub propose a uniform measure of states that
they interpret as apparatus states. They perform the integration directly
for the two dimensional case, and indicate the integration scheme can be
extended to the more dimensional case. Their result, like ours, is the
reproduction of the Born rule. From the perspective of this paper,
Bayes-optimal observation yields an interpretation for the regions employed
by Bohm and Bub.

\section{Concluding remarks}
The search for a Bayesian or decision-theoretic framework for quantum probability has
recently been subject of a number of interesting publications (\cite{Caves}%
, \cite{Deutsch}, \cite{Fuchs}, \cite{Lehrer}, \cite{Pitovski}, \cite%
{Saunders}, \cite{Spekkens}, and \cite{Wallace}). One important motivation
for seeking such an interpretation, is that it allows for a subjective
interpretation of quantum probability by regarding the state vector as a
mathematical representation of the knowledge an agent has about a system.
An often heard critique of Bayesian
interpretations of quantum probability is that, from a strictly Bayesian point of
view, the state vector represents the knowledge available to the
agent that deals with it. A majority of physicists rejects this notion,
mainly because they feel the relative frequencies obtained in actual
experiments are objective features of the system, and not of the knowledge
of the agent. The Bayesian pragmatic response to this, is that what can be inferred
about a system always depends on one's prior knowledge of the system.  
However, in a theory that takes observation as a primitive concept, one
cannot assume to have \emph{a priori} knowledge. This is what we have modelled here as the
uniform distribution of initial observer states and Bayes-optimal observation
of an ensemble of identical states will then result in an unbiased
probability. If it is physically possible to obtain unbiased estimates for a sufficient number of observables so that we can 
reconstruct the state vector, then, at least in an operational sense, the state can be truly assigned in an objective way to
a system. Besides the objective informational content of the state, the state may also represent an objective reality.
This is in agreement with the fact that we started from assumption
(\ref{interaction}); that the state is a realistic description of the
system, and it is the state of the system and the observer that physically
and deterministically interact to produce the measurement outcome. Systems 
\emph{are} in a state, and that state uniquely determines every possible
interaction. The state vector truly represents complete information about a
system, but not merely as a collection of objective attributes, but as a
representation of the possible deterministic interactions with any other
system, in particular observing systems. A classically objective
attribute, from this perspective, is then the limiting
case where the same result follows for the vast majority of states of Bayes-optimal
observing systems that the system can interact with.

The proposed interpretation is falsifiable in principle but there are obstacles along the way. 
If we succeed in preparing the relevant degrees of freedom of the states of the apparatus, we could produce a non-uniform distribution
for the initial states. Such a prepared apparatus would be able to distinguish some pairs
of states better, and some pairs of states worse than the usual Born rule
allows, which means it can only be used to our advantage if
we posses \emph{some} information about the state prior to the measurement.
It also means that the probability for the occurrence of an outcome when we
measure a mixture of states, depends nonlinearly on the probabilities for
each component of the mixture; a failure of what we have called the
\textquotedblleft linear mixture property\textquotedblright . This would
most likely lead to a rejection of the validity of the apparatus by the
majority of experimentalists. And, we hope to have shown, in complete absence of prior information,
it is not evidently desirable to deviate from a complete lack of knowledge of the
apparatus state.

Perhaps there is another, still deeper, reason why it is not possible to
completely control the state of the observer at the quantum level. The source of
probability in observation, the randomness in the state of the observer, may very well at some point become 
\emph{fundamentally} incontrollable. Logical arguments seem to defend at least the possibility of such a thesis. 
In \cite{Breuer1995}, it is shown by an
elegant construction, that for every observer there will be different states
of himself that he cannot distinguish. In \cite{Aerts undecidable} it is
shown that, on purely logical grounds, no observer can determine whether his
observations are entirely faithful\footnote{%
To the best of our knowledge, the relation to such logical arguments and the
quantum measurement problem was pointed out for the first time in 1977 in a
remarkable pioneering paper by Dalla Chiara in \cite{Dalla Chiara 1977}.}.
It seems that, for every single measurement outcome, there is a trade-off between the
information an observer can choose to extract about himself, and about the
system he is observing. This trade-off can be quantified. It is argued in 
\cite{Szilard} and \cite{Alicki 2004} on thermodynamical grounds, that any
gain in information about a system is accompanied by an equal increase of
entropy about the state of the observing system. If this is indeed the
underlying structure for the occurrence of the quantum probabilistic
structure, then the probabilities in quantum mechanics are indeed ontic and
epistemic at the same time. From an absolute perspective, probability always
arises because there is a lack of knowledge situation; it is a measure over
deterministic events. But to the one who observes, this lack of knowledge
may be fundamentally irreducable. It might turn out that, after all, Einstein
and Bohr were both right about the origin of probabilities in quantum
mechanics.
\newline
\newline

{\bf Acknowledgements:} I want to thank Freddy De Ceuninck and Michiel Seevinck for reading and
discussing the content of this paper. This work was supported by the Flemish
Fund for Scientific Research (FWO) project G.0362.03.

\section{Appendices}

\subsection{Appendix A}

\begin{lemma}
If $\mu $ is a (probability) measure such that $\mu (\Delta _{n-1}(X))=1,$
and $C_{k}^{s}$ is defined by the convex closure of (\ref{eigset}), then we
have $\mu ([C_{k}^{s}])=t_{k}$
\end{lemma}

\begin{proof}

Let $\rho _{n-1}$ be the (not necessarily normalized) ($n-1)$ -Lebesgue
measure in $\Delta _{n-1}(X)$. Then we have%
\begin{eqnarray*}
\mu ([C_{k}^{s}]) &=&\frac{\rho _{n-1}([C_{k}^{s}])}{\rho _{n-1}(\Delta
_{n-1})} \\
&=&\frac{\rho _{n-1}([x_{1},\ldots ,x_{k-1},\mathbf{x}(\psi
^{s}),x_{k+1},\ldots ,x_{n}])}{\rho _{n-1}([x_{1},\ldots ,x_{n}])} \\
&=&\frac{\rho _{n-2}\left( B\right) d(B,\mathbf{x}(\psi ^{s}))}{\rho
_{n-2}\left( B\right) d(B,x_{k})}
\end{eqnarray*}%
In this last equation, $B=[x_{1},\ldots ,x_{k-1},x_{k+1},\ldots ,x_{n}]$ is
the face shared by the two simplices, and $d(B,a)$ the smallest Euclidean
distance between point $a$ and each point of face $B,$ which is proportional
to the norm of the orthogonal projection of $a$ onto a unit vector $b$
perpendicular to $B.$ In $\mathbb{R}^{n}$ no unique vector is perpendicular to $B$ (which only has affine
dimension $n-2$), but as long as we stick to the same vector $b$ for both
simplices, the same constant of proportionality will apply, and the ratio
will eliminate that constant. Pick the $x_{k}$ base vector as $b$, which is
obviously unit-norm and perpendicular to $B.$ The orthogonal projection of
the top of $C_{k}^{s}$ to $b$ is: $\mathbf{x}(\psi ^{s})\downarrow b=\langle 
\mathbf{x}(\psi ^{s}),x_{k}\rangle x_{k}=t_{k}x_{k}$ . For $\Delta _{n-1},$
the top is the vector $x_{k}$ itself and its projection $x_{k}\downarrow
b=\langle x_{k},x_{k}\rangle x_{k}=x_{k}.$ Hence we have%
\begin{eqnarray*}
\frac{d(B,\mathbf{x}(\psi ^{s}))}{d(B,x_{k})} &=&\frac{||(\mathbf{x}(\psi
^{s})\downarrow b)||}{||(x_{k}\downarrow b)||} \\
&=&t_{k}x_{k}/x_{k}=t_{k}
\end{eqnarray*}

\end{proof}

\subsection{Appendix B}

\begin{lemma}
Let $C_{k}^{s}$ be defined as in (\ref{eigset}) and $eig(x_{k},\mathbf{x}%
(\psi ^{s}))$ by (\ref{real eigensets}), then: 
\begin{equation*}
C_{k}^{s}\subset \mathbf{\ }eig(x_{k},\mathbf{x}(\psi ^{s}))\subset \lbrack
C_{k}^{s}]
\end{equation*}
\end{lemma}

\begin{proof}
We start with the first inclusion. Suppose $\mathbf{x}(\psi ^{m})$ is in one
of the open $(n-1)-$simplices $C_{k}^{s},$ then, by definition, there exist $%
\lambda _{i}$ such that, with $0<\lambda _{i}<1,\sum \lambda _{i}=1,$ 
\begin{equation}
\mathbf{x}(\psi ^{m})=\sum_{i\neq k}^{n}\lambda _{i}x_{i}+\lambda _{k}%
\mathbf{x}(\psi ^{s})  \label{s}
\end{equation}%
On the other hand we have that $\mathbf{x}(\psi ^{s})\in \Delta _{n-1}$ and
hence there exist $t_{l}\geq 0,\sum t_{l}=1$ such that (\ref{stat state})
holds: 
\begin{equation}
\mathbf{x}(\psi ^{s})=\sum_{l=1}^{n}t_{l}x_{l}  \label{a}
\end{equation}%
Substitution of (\ref{a}) into (\ref{s}) yields%
\begin{equation*}
\mathbf{x}(\psi ^{m})=\lambda _{k}t_{k}x_{k}+\sum_{i\neq k}^{n}(\lambda
_{i}+\lambda _{k}t_{i})x_{i}
\end{equation*}%
Calculating the likelihood ratios (\ref{odds}), we obtain $\Lambda _{k}=%
\frac{1}{\lambda _{k}},$ and for $i\neq k$ we have:%
\begin{equation*}
\Lambda _{i}=\frac{t_{i}}{\lambda _{i}+\lambda _{k}t_{i}}
\end{equation*}%
We easily see that $\Lambda _{k}>\Lambda _{i}$ iff $\lambda _{i}>0,$ which
is satisfied by assumption. Hence, by (\ref{real eigensets}) every $\mathbf{x%
}(\psi ^{m})\in $ $C_{k}^{s}$ gives an outcome $x_{k}$, establishing the
result. For the second inclusion, suppose there exists some $\mathbf{x}(\psi
^{m})\in \Delta _{n-1}$ with $\mathbf{x}(\psi ^{m})\notin \lbrack
C_{k}^{s}]. $ The sets $C_{k}^{s}$ in our theorem, as can be seen from the
definition (\ref{eigset}), are disjoint open $(n-1)$-simplices. If we had
defined them by means of the \emph{closed} convex closure, they would
maximally share the $(n-2)$ simplex $\Delta _{n-2}^{s}(j,k)=[\mathbf{x}(\psi
^{s}),x_{1},\ldots ,x_{j-1},x_{j+1},\ldots ,x_{k-1},x_{k+1},\ldots ,x_{n}]:$%
\begin{equation*}
\lbrack C_{j}^{s}]\cap \lbrack C_{k}^{s}]=\Delta _{n-2}^{s}(j,k)
\end{equation*}%
Assume first $a$ is not in the boundary of $[C_{k}^{s}],$ i.e. not in one of
the lower dimensional sub-simplices $\Delta _{n-2}^{s}(j,k).$ Then $\mathbf{x%
}(\psi ^{m})$ $\in C_{i}^{s}$ with $i\neq k.$ Because of the above
demonstrated first inclusion we have $\mathbf{x}(\psi ^{m})\in eig_{\mathbf{x%
}(\psi ^{s})}(x_{i})$ and hence $\mathbf{x}(\psi ^{m})\notin eig(x_{k},%
\mathbf{x}(\psi ^{s})).$ If on the other hand $\mathbf{x}(\psi ^{m})\in
\Delta _{n-2}^{s}(j,k)$, our outcome assignment on the basis of the maximum
likelihood principle is ambiguous, as there will be two equal maxima, and
even more when $\mathbf{x}(\psi ^{m})$ is chosen in a still lower
dimensional subsimplex. However, we are free to choose whatever outcome we
like as long as it is one of the maxima. Because the maxima coincide, these
points lie in the boundary and hence the conclusion remains $eig(x_{k},%
\mathbf{x}(\psi ^{s}))\subset \lbrack C_{k}^{s}].$
\end{proof}

\subsection{Appendix C}

\begin{lemma}
The mapping $\omega $%
\begin{eqnarray*}
\omega &:&S_{n}\rightarrow \Delta _{n-1} \\
\omega (z) &=&(z_{1}z_{1}^{\ast },z_{2}z_{2}^{\ast },\ldots
,z_{n}z_{n}^{\ast })
\end{eqnarray*}%
is measure-preserving, i.e. for two measure spaces $(\Delta _{n-1},{\mathcal{%
B}}(\Delta _{n-1}),\mu )$ and $(S_{n},{\mathcal{B}}(S_{n}),\nu )$ and $A\in $
${\mathcal{B}}(\Delta _{n-1})$ and $\omega ^{-1}(A)\in {\mathcal{B}}(S_{n})$%
, we have:

\begin{equation*}
\nu (\omega ^{-1}(A))=\frac{2\pi ^{n}}{\sqrt{n}}\mu (A)
\end{equation*}
\end{lemma}

\begin{proof}
\footnote{%
This proof was first presented in \cite{Aerts Born}, but we include it for
completeness. The author is grateful for a valuable hint from Wade Ramey
that was helpful in proving the theorem.}Let $A$ be an arbitrary open convex
set in $\Delta _{1}:A=\{(x_{1},x_{2}):a<x_{1}<b,\ x_{2}=1-x_{1}\}$.
Evidently, $\mu (A)=\sqrt{2}(b-a)$. Let $B$ be the pull-back of $A$ under $%
\omega :$ 
\begin{eqnarray*}
B &=&\{(z_{1},z_{2})\in Z_{1}\times Z_{2}\subset \mathbb{C}%
^{2}:Z_{1}=\{z_{1}:a<|z_{1}|^{2}<b\}, \\
Z_{2} &=&\{z_{2}:\ z_{2}=\sqrt{1-|z_{1}|^{2}}e^{i\theta },\theta \in \lbrack
0,2\pi \lbrack \}\}
\end{eqnarray*}%
Clearly, 
\begin{equation*}
\nu (B)=\nu (Z_{1})\nu (Z_{2})=\pi (b-a).2\pi =\frac{2\pi ^{2}}{\sqrt{2}}\mu
(A)
\end{equation*}%
Hence the theorem holds for convex sets in $\Delta _{2}$. This conclusion
can readily be extended to an arbitrary $(n-1)$-dimensional rectangle set $A$
in $\Delta _{n-1}:$%
\begin{equation*}
A=\{(x_{1},\ldots ,x_{n-1},1-\sum_{i=1}^{n-1}x_{i}):\forall i=1,\ldots
,n-1:a_{i}<x_{i}<b_{i};\ a_{i},b_{i}\in \lbrack 0,1]\}
\end{equation*}%
Its measure factorizes into:%
\begin{equation*}
\mu (A)=\sqrt{n}\prod_{i=1}^{n-1}(b_{i}-a_{i})
\end{equation*}%
Next consider n-tuples of complex numbers: 
\begin{eqnarray*}
B &=&\{(z_{1},z_{2},\ldots ,z_{n})\in Z_{1}\times \ldots \times Z_{n}\} \\
Z_{i} &=&\{z_{i}\in \mathbb{C}:a_{i}<|z_{i}|^{2}<b_{i},i\neq n\},\  \\
z_{n} &=&\sqrt{1-|z_{1}|^{2}-\ldots -|z_{n-1}|^{2}}e^{i\theta _{n}},\theta
_{n}\in \lbrack 0,2\pi \lbrack \}\}
\end{eqnarray*}%
Clearly $\omega (B)=A$. The measure of $B$ can be factorized as:%
\begin{eqnarray*}
\nu (B) &=&\nu (Z_{1})\nu (Z_{2})\ldots \nu (Z_{n}) \\
&=&2\pi \prod_{i=1}^{n-1}\pi (b_{i}-a_{i})=\frac{2\pi ^{n}}{\sqrt{n}}\mu (A)
\end{eqnarray*}%
Hence the theorem holds for an arbitrary rectangle set $A\subset \Delta
_{n-1}.$ But every open set in $\Delta _{n-1}$ can be written as a pair-wise
disjoint countable union of rectangular sets. It follows that $\nu (\omega
^{-1}(\cdot ))=\frac{2\pi ^{n}}{\sqrt{n}}\mu (\cdot )$ for all open sets in $%
\Delta _{n-1}$. Both $\nu $ and $\mu $ are finite Borel measures because $%
\Delta _{n-1}$ and $S_{n}$ are both compact subsets of a vector space of
countable dimension. Therefore they must be regular measures (\cite{Rudin
1987}, p47), which are completely defined by their behavior on open sets.
Hence $\omega $ is measure preserving for Borel sets.
\end{proof}
\end{document}